# Orbiting of bacteria around micrometer-sized particles entrapping shallow tents of fluids

George Araujo[1], Weijie Chen[1,2], Sridhar Mani[2], Jay X. Tang[1*]

[1] Department of Physics, Brown University, 182 Hope st, Providence, RI, 02912, USA

[2] Department of Genetics, Albert Einstein College of Medicine, 1300 Morris Park Avenue, Bronx, NY 10461, USA

*Corresponding author: Jay_Tang@brown.edu

**Abstract**

**Hydrodynamics and confinement dominate bacterial mobility near solid or air-water boundaries, causing flagellated bacteria to move in circular trajectories. This phenomenon results from the counter-rotation between the bacterial body and flagella and lateral drags on them in opposite directions due to their proximity to the boundaries. Numerous experimental techniques have been developed to confine and maneuver motile bacteria. Here, we report observations on *Escherichia coli* and *Enterobacter sp.* when they are confined within a thin layer of water around dispersed micrometer-sized particles sprinkled over a semi-solid agar gel. In this setting, the flagellated bacteria orbit around the dispersed particles akin to planetary systems. The liquid layer is shaped like a shallow tent with its height at the center set by the seeding particle and the meniscus profile set by the strong surface tension of water. The tent-shaped constraint and the left handedness of the flagellar filaments result in exclusively clockwise circular trajectories. The thin fluid layer is resilient due to a balance between evaporation and reinforcing fluid pumped out of the agar. The latter is driven by the Laplace pressure caused by the curved meniscus. This novel mechanism to entrap bacteria within a minimal volume of fluid is relevant to near surface bacterial accumulation, adhesion, biofilm growth, development of bio-microdevices, and cleansing hygiene.**

**Significance**

We report a novel observation of thin but robust tent-like layer of liquid around micro-sized particles sprinkled over a semisolid agar surface. The layer is resilient to evaporation and provides a strong confinement that allows orbital motion of flagellated bacteria around the particles. The experimental setup is simple and it is surprising that such small particles each retain a water layer, sustaining circular bacterial trajectories. Our discovery is explained by applying a hydrodynamic model on bacterial swimming. The mechanism is relevant to practical situations involving bacterial behavior in environments with scarcity of water, biofilm growth, infection control, etc.

**Introduction**

A flagellated bacterium propels itself by rotating its flagella relative to its body, which counterrotates [1]. Freely swimming bacterial cells tend to accumulate close to solid surfaces [2-4], with their residence times within a few micrometers (up to 1-3 μm) from the boundary lasting several seconds due to hydrodynamic effects [5]. The coupling of body rotation and the enhancement in the drag force exerted on the part of the cell closer to the solid surface causes the bacterium to move in circular trajectories [6]. The handedness of the flagella determines whether the circular trajectory is clockwise or counterclockwise. Propelled by its flagellar filaments with left-handed helical structure, *E. coli* moves forward in clockwise paths when looked from above [7]; it swims on the right-hand side when prevented to turn by a solid wall on its side [8].

Numerous experimental techniques, including microfluidic devices [9,10], droplets [11-14], or vortices [15], have been developed to confine and maneuver motile bacteria. The mode of bacterial motility can also be shaped by confinement. For example, whereas flagellated bacteria swim in liquid, on a solid surface they may move collectively in the form of swarming motility [16-18]. The swarming motility triggers a variety of adhesive mechanisms, including those preceding to the growth of biofilms [19-21].

**Results**

Our experiment starts by allowing swarming bacteria to break away into swimming by adding a tiny droplet of water (2-5 μl) next to a swarming colony (see Method; illustrated in Fig 1a). A number of bacteria leave their pack at the swarm front and disperse over the agar surface (Fig. 1b). As the droplet of water evaporates, most bacteria get stuck to the surface and cease the swimming motion within a short time interval (~1 minute). However, graphite particles or silica beads contained in the water droplet entrap a layer of water in their surrounding regions, allowing a few bacteria to swim around the particles, exclusively in clockwise trajectories (Fig. 1c&1d). Such orbiting motion lasts over 10 min during observation, much longer time than expected for a water film a few micrometers in thickness to disappear due to evaporation.

We observed common orbiting motion among a wild type strain of *E. coli* (HCB33) and three recently isolated strains of bacteria, including two novel enterobacter isolates, SM1 and SM3, and one *E. coli* isolate, H5 (details in the Methods). For the two *E. coli* strains, we used agar containing added surfactant Triton X-100 (Sigma-Aldrich) of 0.1% vol/vol in order to facilitate the colony spread. We also experimented with *Pseudomonas aeruginosa* (PA01) and *Bacillus subtilis* (3610). In these two latter cases, we also noticed the long-lasting water layers around the micrometer-sized particles where bacteria were motile. For *P. aeruginosa*, we observed orbiting around the micro-sized particle, but the motion was less organized, i.e. bacteria tended to go off track from their orbits within a few seconds, possibly due to this species of bacteria frequently reversing flagella motor rotation direction [22]. *B. subtilis*, which are longer cells (4-10 μm in length) remained motile within the water meniscus, but they did not orbit well. These results suggest universal motion of bacteria entrapped around microsized particles on moist surface, although the exact trajectories are species dependent. From here on, we focus the remaining report on *Enterobacter sp.* and *E. coli*, which reliably form circular trajectories.

Small graphite particles (< 2.5 μm in size) are usually not surrounded by swimming cells, possibly because they could not retain the water reservoir large enough to entrap bacteria. Large graphite particles (> 12 μm in size) retain liquid in a region over 40 μm in radius and thus keep many bacteria swimming in their proximity. Intermediate graphite particles were frequently seen to be surrounded by a few active swimmers and were the ones around which the orbiting of bacteria was most often observed and measured (Fig. 1d). In cases where the number of swimmers around a dispersed particle is big (over a dozen, as is commonly seen around large graphite particles or particles immersed in a dense bacterial population), the cells frequently bump into each other, making it difficult to discern their trajectories. We chose silica beads of uniform size (4.5 μm diameter; Bangs Laboratories, Inc.) to perform the same experiment and observed equivalent results between the beads (Fig. 2a) and the graphite particles of comparable size, suggesting that the key parameter is the particle size, not its chemical nature. The common role of the graphite particles or beads is to sustain a tent-shaped water reservoir around them.

Around the particles where bacteria were observed to orbit, the radii of meniscus of water and bacterial trajectories correlate to the particle size. For the silica beads, which are uniform in dimensions, the meniscus and trajectory radii vary within narrow ranges dependent on the extent of evaporation. Graphite particles, which were more variable in size and had irregular shapes, gave rise to larger variation in meniscus and trajectory radii (Fig. 2b&2c).

**Discussion**

**Hydrodynamic cause for the circular trajectories.**
The circular trajectories are caused primarily by the solid confinement imposed by the agar surface. The orbiting bacteria were observed to be in the same focal plane as the bacterial cells stuck in the nearby dry regions. The surface contour of graphite particles had minimal effects on the bacterial trajectories with exceptions of close encounters. In some cases bacteria were seen to swim within a body length from the center particle, but often orbiting bacteria swam over 8 μm away from it, a distance by which the hydrodynamic effect of the graphite boundary becomes negligible [5]. These observations reinforce the argument that the circular trajectories are caused by the proximity of the bacteria to the bottom surface. The observation that *Enterobacter sp.* move in the same clockwise direction as *E. coli* allows us to conclude that both species of bacteria have left-handed helical flagella.

Measurements on all four strains show a relationship between the speed of the orbiting bacteria and the radii of curvature of their trajectories. Bacteria swam slower in trajectories of smaller radii, but beyond a certain threshold (~4 μm) the speed approached an upper limit, between 30-50 μm/s, depending on the bacterial strain (Fig. 3a). These results indicate that graphite particles or beads only affect the swimming speed by restricting the bacteria to their close proximity via the tent-shaped liquid film around them.

The experimental speed profile agrees with a hydrodynamic model described previously [23] (details on its implementation discussed in Methods). Although it is possible to shift the theoretical line (Fig. 3a), by adjusting the parameters for the dimensions of the model bacterium and the flagellar motor rotation speed, the key feature remains the same: the speed initially increases with the radius of curvature, but quickly attains an upper value. The consistent trend of the data provides further support that the circular motion observed in the orbiting bacteria is mainly caused by the hydrodynamic boundary effect of the bottom surface (Fig 3b).

**Endurance of the water layer around the seeding particle.**

One key finding of our study is that graphite or bead particles on the agar surface retain a layer of water around them much longer than on a non-permeable surface. The evaporation of a water layer from a borosilicate glass surface containing microspheres on top has been investigated experimentally using x-ray transmission microscopy [24] (polystyrene beads of 6 μm diameter). The full process lasts about 12 seconds, with 96% of the time on a pinning stage when the water film surrounding the bead thins gradually, and the last 4% on depinning of the water meniscus around the microsphere in the form of a sudden collapse. In our case, however, the fact that the bottom surface is made by agar gel (>97.5% water) allows for water to constantly permeate the upper surface. Thus, the meniscus lasts much longer (Figure 4). This observation is explained as a balance between evaporation and additional water pumped from the agar gel to the layer. Given the tiny thickness (a few μm) of the water layer, its internal pressure P' is roughly constant everywhere. The internal pressure within the water layer (P') is lower than the atmospheric pressure $P_o$ [25]. The difference between these two pressures is given by the Young-Laplace equation, $P_o - P' = \sigma/R$, where σ is the water surface tension and $1/R$ is the meniscus total curvature. Near the top of the agar gel, the pressure is also close to the atmospheric pressure. Hence, a pressure gradient is established between the permeable agar gel and the water layer, giving rise to water being pumped, as expected from Darcy's law [26]. Consequently, even though water is constantly evaporating from the thin layer, it is being replenished by that from the agar gel.

**Conclusion**

In summary, we report a novel mechanism by which micro-sized particles on a moist surface entrap a tent-shaped layer of liquid within which flagellated bacteria swim in circular trajectories. The phenomenon occurs due to strong surface tension at the liquid-air interface and the proximity of the cells to a semi-permeable agar gel. Our finding implicates the potential for persistence of bacteria in environments with restricted water supply. For instance, certain bacteria cooperate with plants to maintain their growth and longevity through droughts [27]. The observation that small particles in a wet environment can entrap bacteria and allow them to remain viable and motile, even after most of the region has dried, raises awareness on the importance of proper hygiene habits to prevent bacterial infection [28,29].

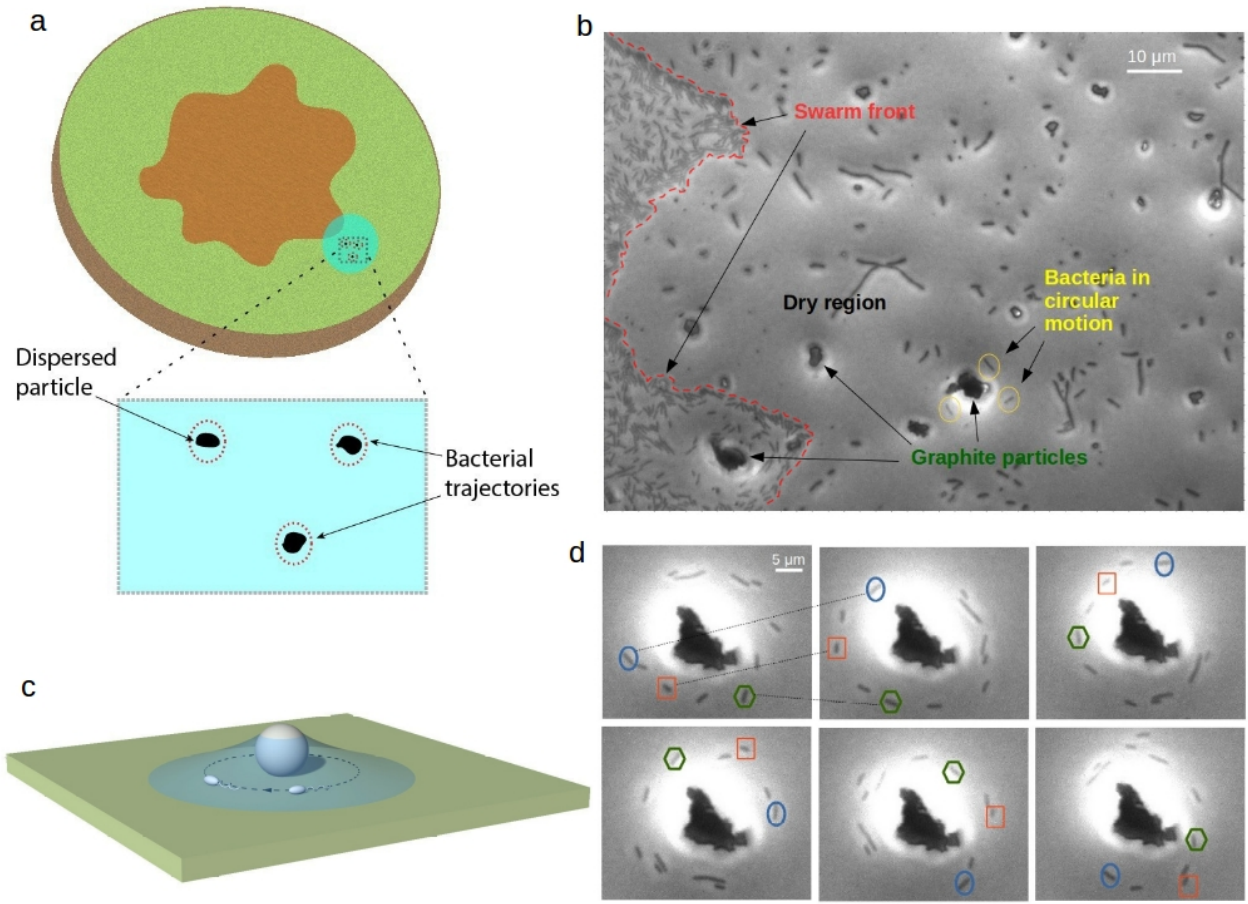

**Fig. 1. Circular motion of bacteria around particles on agar surface.** (a) Schematics representing a bacterial swarm (in purple) with a water droplet (containing graphite particles or silica beads) added to a location overlapping the swarm front. A region in the droplet is amplified to illustrate bacterial trajectories around the dispersed particles. (b) Microphotograph showing different features following the water droplet evaporation. The swarm front is marked by the red dashed line. Most of the region beyond has dried and is populated by bacteria stuck to the surface, except for those around graphite particles. (c) Artistic representation of bacterial cells moving around a dispersed particle within a tent-like layer of water around it. (d) Time lapse following the trajectories of bacterial cells orbiting a graphite particle. Three cells in particular are highlighted with different colors and geometrical contours to aid visualization. The sequence goes from left to right and the pictures shown were taken approximately 0.47 s apart. These images were extracted from Supplementary Movie 1.

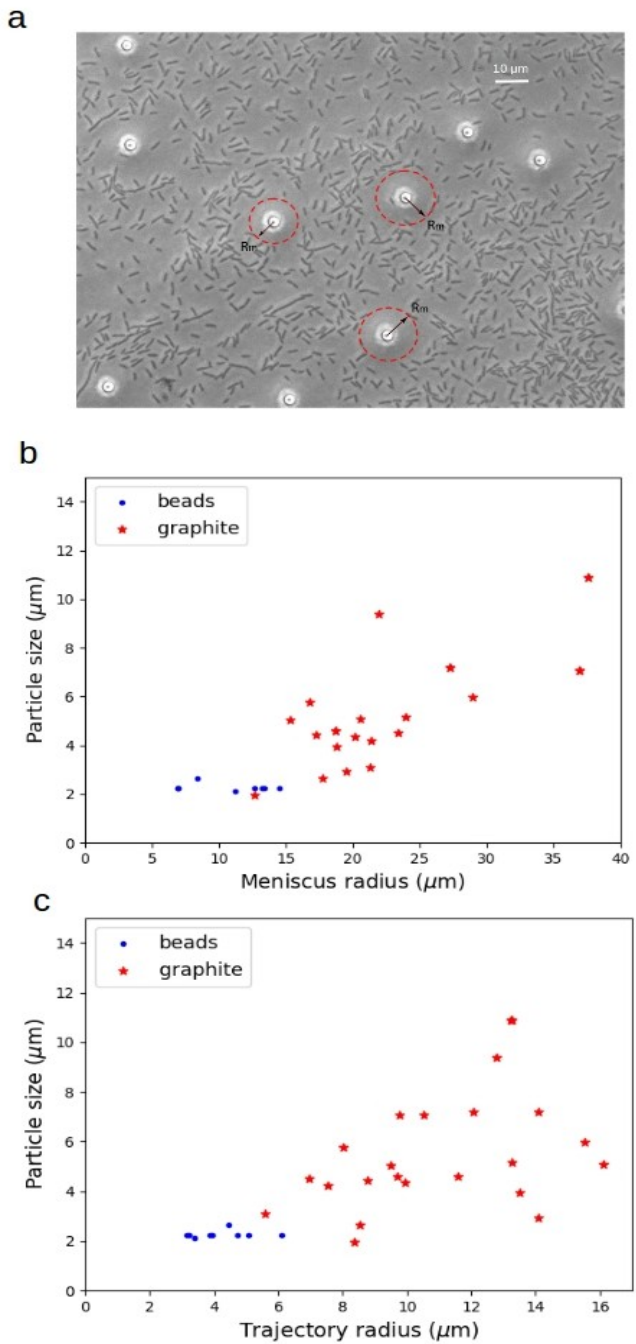

**Fig. 2. Image of meniscus of water layer and the measured dependence of both meniscus and trajectory radii on particle size.** (a) Microphotograph highlighting the meniscus edge of water (radius $R_m$) around silica beads. Bacteria outside these circles are stuck in dried regions, whereas a few bacteria within these circles orbit around the beads (Supplementary Movie 2). (b) Relationship between the size of dispersed particles and the cutoff radius of their surrounding water meniscus. Each data point represents an individual observation. (c) Relationship between the size of the dispersed particles (graphite or beads) and the radius of trajectories of the orbiting bacteria.

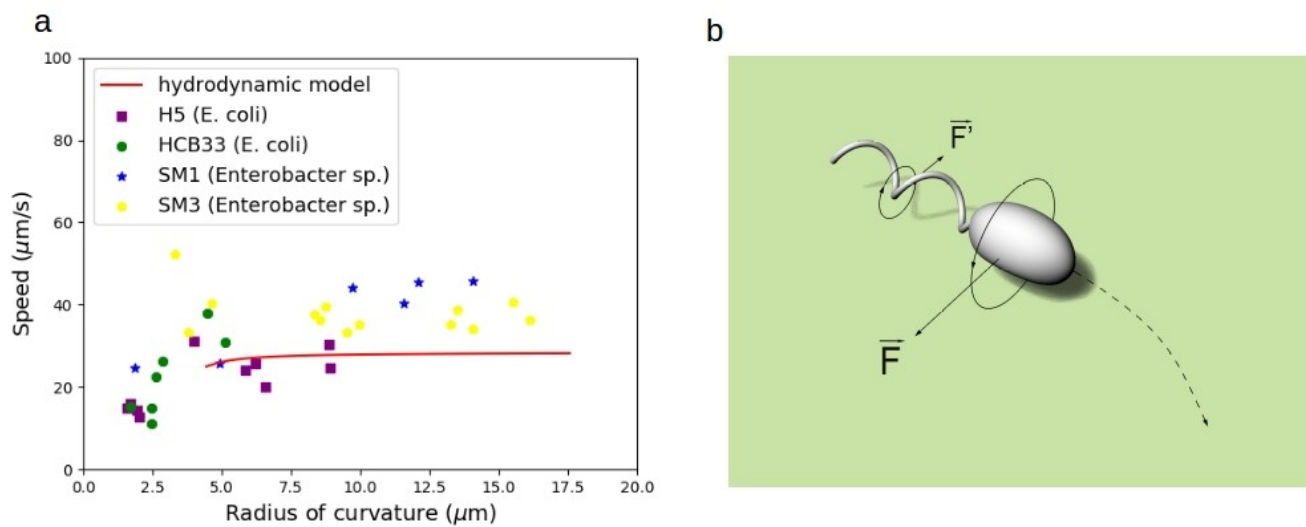

**Fig. 3. Measured speed versus radius of trajectory in comparison with numeric prediction by applying a simple hydrodynamic model**. (a) Speed of bacteria with respect to the radius of curvature of trajectories for the four strains tested. The red line is the prediction by applying the Resistive Force Theory (with parameters specified in the Methods). The lower end of the line is set by the 5 nm minimal gap chosen between the cell body and the bottom surface in the model. Each data point represents a measurement for an individual bacterium. (b) Schematics showing the curvilinear trajectory of a bacterium when it swims close to a solid boundary. The cell body and its flagellar bundle rotate in opposite directions. The larger drag on the cell body at the location facing the surface exerts a lateral force in the opposite direction to that on the flagella (forces indicated in the figure). The net torque leads to a circular trajectory.

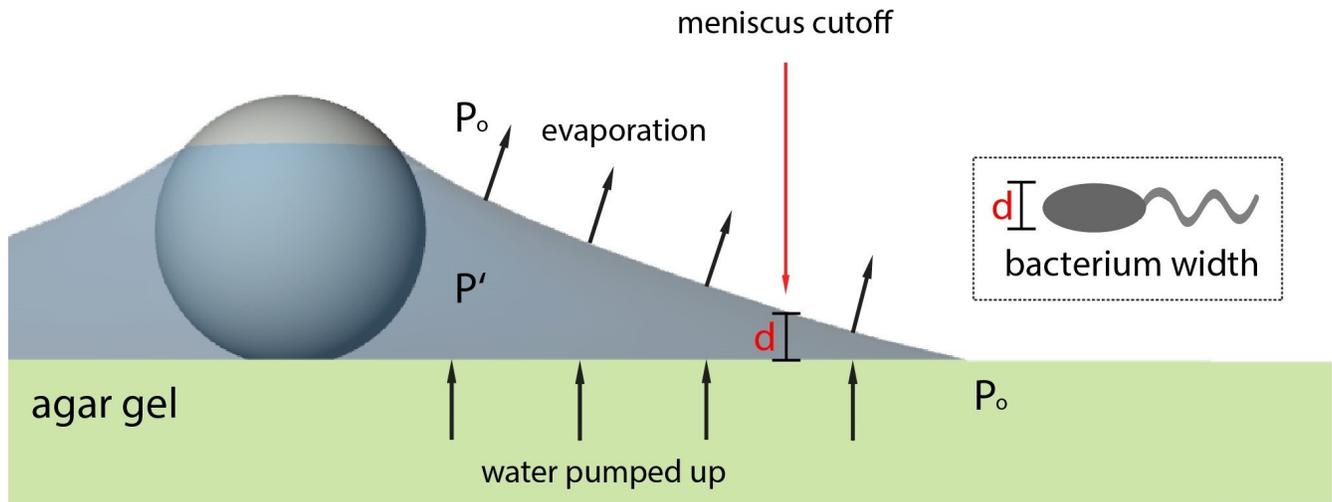

**Fig. 4. Schematics illustrating a resilient water layer around a micron-sized particle on an agar gel.** The internal pressure (P') within the layer of water is lower than the atmospheric pressure ($P_o$), which is also roughly the pressure inside the agar gel close to the top surface. This pressure difference pumps water up from the semisolid agar gel, counterbalancing constant evaporation. Hence, the water layer is maintained much longer than on typical non-permeable solid surfaces. The thickness d corresponds to the width of a bacterial cell, with its location defining a meniscus boundary outside which bacteria get stuck to the agar surface.

## Materials and Methods

### Bacterial strains

Six strains of bacteria were used. Strain HCB33 (wild type E. coli) was obtained from Howard Berg (Harvard University). *P. Aeruginosa* PA01 was obtained from Keiko Tarquinio (Emory University Medical School). *B. subtillis* 3610 was obtained from Daniel B. Kearns (Indiana University). The remaining three strains were isolated at the Albert Einstein College of Medicine (NYC). H5, a swarming strain of *E. coli* from human feces (Committee on Clinical Investigation CCI# 2009-446-006; IRB# 2015-4465). SM1 and SM3 are novel strains of swarming Enterobacter sp. isolated from mouse feces (presently undergoing species characterization).

### Bacterial culture

Bacteria were grown in suspensions of Lysogeny Broth (LB) medium overnight (~ 16 hours) and then a 5 μl droplet was taken to inoculate the center of a petri dish containing LB agar gel (LB + 0.5% agar). Triton-X100 (Sigma-Aldrich) was added (0.1% in volume) to the LB agar as external surfactant for *E. coli* inoculation in order to help with its colony spread. In particular, it was observed for the H5 strain that the colony spreads in similar rate to what is seen in typical swarming colonies: it populates the whole petri dish (~10 cm in diameter) within a few hours. For the HCB33, the effect was not as dramatic and the colony grew only slightly faster compared to situations when no external surfactant is added to the growth medium.

The samples were placed into an incubator with controlled temperature of 37 ºC and 50% relative humidity for 3 to 5 hours, depending on the growth rate. In the next step, a droplet containing micrometer-sized particles (graphite or silica beads) was placed at a point overlapping the colony, as illustrated in Figure 1a. Graphite particles were obtained by scrapping a pencil with a razor blade and dispersed in deionized water (DI water). Silica beads (4.5 μm in diameter, Bangs Laboratories, Inc) were also diluted in DI water. After the addition of the water droplet, the petri dish was gently tilted in different directions to help the droplet to spread away from the bacterial colony. This motion created thin water layers that evaporated within one minute on the agar surface. Most bacteria then got stuck on the dried surface, leading a small number of swimming bacteria trapped in water reservoirs around the dispersed particles.

### Imaging and measurements

The droplet region was imaged under an upright microscope (Nikon Eclipse E800) with a 40x objective lens, coupled to a CCD camera (Photometrics Coolsnap EZ). The imaging was controlled by the software Metamorph (Molecular Devices) from which the movies were recorded. The recording rate was in the range between 10.72 frames per second (fps) to 33.88 fps, with lower frame rates chosen for bigger imaging area.

Radius of meniscus was determined by defining the nearby immobilized bacteria as the cutoff perimeter, as indicated on Figure 2a. Bacterial trajectories and their swimming speeds were calculated using ImageJ from their captured positions in video clips. The numerical calculation applying the Resistive Force model was implemented in a Python code.

**Hydrodynamic model**

E. Lauga *et. al* proposed a model based on Resistive Force theory to represent bacteria moving close to solid boundaries [23]. It models a bacterium as formed by a sphere connected to a helical rod, to represent the cell body and flagella bundle, respectively. The free swimming organism is subjected to zero net force and torque and is propelled by the rotating flagella bundle. Body and flagella are characterized by their respective mobility matrices which appear in the linear system of equations, equation (1), given the velocities and rotation rates of the cell body in equations (2) and (3), respectively.

$$(A+B)X = BY \qquad (1)$$

where,

$$X = (U_x, U_y, U_z, \Omega_x, \Omega_y, \Omega_z)^T \qquad (2)$$

$$Y = (0, 0, 0, 0, \omega, 0)^T \qquad (3)$$

Matrix A is the mobility matrix of the cell body. B is the mobility matrix of the flagella. $U_i$ gives the i-th component of the velocity and $\Omega_i$ the i-th component of the rotation rate of the cell body. The rotation rate of the flagella is given by $\omega$. The cell is set to initially align with the y axis, without loss of generality. Reference [23] gives the details on how to build the mobility matrices.

The velocities and rotation rates in this model are dependent on the gap distances between the surface and the material parts of the model bacterium. The speed U of the circular moving bacterium is given by the components along the bottom surface: $U = (U_x^2 + U_y^2)^{1/2}$ and the radius of curvature R of the cell trajectory is also obtained from the solution of the linear system: $R = U/|\Omega_z|$.

Acknowledgements: We acknowledge funding from NSF DMR 1505878 (JXT), NIH R01 CA CA161879 (SM), and CAPES BEX 13733-133 (GA). We thank Neha Mani for the initial idea of her school project that led to the experimental design. We also thank Yinan Liu for her assistance on graphic illustration.

Author Contributions: GA, WC & JXT conceived the study. GA and WC performed the experiments and analysis. SM produced and provided the bacteria (SM1, SM3 and H5). GA and JXT wrote the paper with contributions from all co-authors.

# References

[1] Lauga E (2016) Bacterial hydrodynamics. Annu Rev Fluid Mech 48, 105–130.

[2] Li G, et al. (2011) Accumulation of swimming bacteria near a solid surface. Phys Rev E 84 041932.

[3] Berke AP, Turner L, Berg HC, Lauga E (2008). Hydrodynamic attraction of swimming microorganisms by surfaces. Phys Rev Lett 101 038102.

[4] Li G, Tang JX (2008) Accumulation of microswimmers near a surface mediated by collision and rotational brownian motion. Phys Rev Lett 103 078101 .

[5] Drescher K, Dunkel J, Cisneros LH, Ganguly S, Goldstein RE (2011) Fluid dynamics and noise in bacterial cell-cell and cell-surface scattering. Proc Natl Acad Sci 108 10940-10945 .

[6] Frymier PD, Ford RM, Berg HC, Cummings PT (1995) Three-dimensional tracking of motile bacteria near a solid planar surface. Proc Natl Acad Sci 92 6195-6199 .

[7] Vigeant MA, Ford RM (1997) Interactions between motile Escherichia coli and glass in media with various ionic strengths, as observed with a three-dimensional-tracking microscope. Appl Environ Microbiol 63 3474–3479.

[8] DiLuzio, WR et al. (2005) Escherichia coli swim on the right-hand side. Nature 435 1271–1274 .

[9] Männik J, Driessen R, Galajda P, Keymer JE, Dekker C (2009) Bacterial growth and motility in sub-micron constrictions. Proc Natl Acad Sci 106 14861-14866.

[10] Biondi SA, Quinn JA, Goldfine H (1998) Random motility of swimming bacteria in restricted geometries. AICHE J. 44 1923-1929.

[11] Hennes M, Tailleur J, Charron G, Daerr A (2017) Active depinning of bacterial droplets: the collective surfing of Bacillus subtilis. Proc Natl Acad Sci 114 5958-5963 .

[12] Sokolov A, Rubio LD, Brady JF, Aranson IS (2018) Instability of expanding bacterial droplets. Nature Comm 9 1322.

[13] Wioland H, Woodhouse FG, Dunkel J, Kessler JO, Goldstein, RE (2013) Confinement stabilizes a bacterial suspension into a spiral vortex. Phys Rev Lett 110 268102.

[14] Tuval I, et al. (2005) Bacterial swimming and oxygen transport near contact lines. Proc Natl Acad Sci 102 2277–2282.

[15] Sokolov A, Aranson IS (2016) Rapid expulsion of microswimmers by a vortical flow. Nature Comm 7 11114.

[16] Turner L, Zhang R, Darnton NC, Berg HC (2010) Visualization of flagella during bacterial swarming. Journal of Bact 192 3259-3267.

[17] Zhang HP, Be'er A, Florin EL, Swinney HL (2010) Collective motion and density fluctuations in bacterial colonies. Proc Natl Acad Sci 107 13626-13630.

[18] Copeland MF, Weibel DB (2009) Bacterial swarming: a model system for studying dynamic self-assembly. Soft Matter 5 1174-1187.

[19] Dunne Jr WM Bacterial adhesion: seen any good biofilms lately? (2002) Clin Microbio Rev 15 155-166.

[20] Jin F, Conrad JC, Gibiansky MS, Wong GCL (2011) Bacteria use type-IV pili to slingshot on surfaces. Proc Natl Acad Sci 108 12617-12622.

[21] O'Toole GA, Kolter R (1998) Flagellar and twitching motility are necessary for Pseudomonas aeruginosa biofilm development. Mol Microbiol 30:295-304.

[22] Cai Q, Li Z, Ouyang Q, Luo C, Gordon VD (2016) Singly flagellated Pseudomonas aeruginosa chemotaxes efficiently by unbiased motor regulation. MBio. 7 e00013-1.

[23] Lauga E, DiLuzio WR, Whitesides GM, Stone HA (2006) Swimming in circles: motion of bacteria near solid boundaries. Biophys Journal 90 400–412.

[24] Cho K, et al. (2016) Low internal pressure in femtoliter water capillary bridges reduces evaporation rates. Scientific Reports 6  22232.

[25] Honschoten JW, Brunets N, Tas NR  (2010) Capillary at the nanoscale. Chem Soc Rev, 39 1096–1114.

[26] Cunningham R (1980). Diffusion in Gases and Porous Media. Springer US.

[27] Xu L, et al. (2018) Drought delays development of the sorghum root microbiome and enriches for monoderm bacteria. Proc Natl Acad Sci 115 E4284-E4293.

[28] Grice EA, Segre JA (2011) The skin microbiome. Nat Rev Microbiol 9 244-253.

[29] Ottemann KM, Miller JF (1997) Roles for motility in bacterial host interactions. Mol Microbiol 24 1109-1117.